\documentclass[useAMS,usenatbib]{mn2e}

\usepackage{graphicx}

\title[Enhanced Core Formation Rate by Self-gravity]{Enhanced Core Formation Rate in a Turbulent Cloud by Self-gravity}
\author[Wankee Cho and Jongsoo Kim]{Wankee Cho$^{1}$, and Jongsoo Kim$^{2,3}$\thanks{E-mail:jskim@kasi.re.kr}\\
$^1$Department of Physics and Astronomy, Seoul National University,
Seoul 151-742, Republic of Korea\\
$^2$Korea Astronomy and Space Science Institute, 61-1, Hwaam-Dong, Yuseong-Gu, Daejeon 305-348, Republic of Korea\\
$^3$Astrophysics Group, Cavendish Laboratory, Cambridge University, JJ Thompson Avenue, Cambridge CB3 0HE, The UK}

\begin{document}
\date{Accepted 1988 December 15. Received 1988 December 14; in original form 1988 October 11}

\pagerange{\pageref{firstpage}--\pageref{lastpage}} \pubyear{2002}

\maketitle

\label{firstpage}

\begin{abstract}
We performed a numerical experiment designed for core formation in a self-gravitating, magnetically supercritical, supersonically turbulent, isothermal cloud.  A density probability distribution function (PDF) averaged over a converged turbulent state before turning self-gravity on is well-fitted with a lognormal distribution.  However, after turning self-gravity on, the volume fractions of density PDFs at a high density tail, compared with the lognormal distribution, increase as time goes on.  In order to see
the effect of self-gravity on core formation rates, we compared
the core formation rate per free-fall time (CFR$_{\rm ff}$) from the
theory based on the lognormal distribution and the one from our
numerical experiment.  For our fiducial value of a critical density, 100, normalised with an initial value, the latter CFR$_{\rm ff}$ is about 30 times larger the former one.  Therefore, self-gravity plays an important role in significantly increasing CFR$_{\rm ff}$.  This result implies that core (star) formation rates or core (stellar) mass functions predicted from theories based on the lognormal density PDF need some modifications.  Our result of the increased volume fraction of density PDFs after turning self-gravity on is consistent with power-law like tails commonly observed at higher ends of visual extinction PDFs of active star-forming clouds.
\end{abstract}
\begin{keywords}
ISM: clouds -- methods: numerical -- MHD -- stars: formation -- turbulence.
\end{keywords}

\section{Introduction}

Probability distribution functions (PDFs) for volume and column density fields calculated from turbulence simulations with the isothermal equation of state have been successfully fitted with a lognormal distribution \citep{vaz94, pad97, pas98, nor99, ost99, ost01, vaz01, kri07, lem08, fed08a}. In addition to these numerical works, it has been recently reported that the H$\alpha$ emission measure for the warm ionized medium \citep{hil08} and the densities of the diffuse ionized gas and diffuse atomic gas \citep{ber08} follow the lognormal distribution. 
The lognormal density PDF has also become one of important ingredients of star formation theories based on turbulence.  The analytical models of \citet{pad02} and \citet{hen08} predicted stellar initial or core mass functions, whose mass distributions at lower masses are largely determined by the lognormal density PDF.  \citet{elm02,elm08} used the lognormal density PDF in order to calculate the cumulative mass fraction above a critical number density ($>10^5~{\rm cm}^{-3}$), which is directly related to the star formation efficiency.  \citet{kru05} also measured dimensionless core (star) formation rate per free-fall time (CFR$_{\rm ff}$) based on the lognormal density PDF.  Instead of picking up a specific value of the critical density as in \citet{elm02, elm08}, they determined it by equating the local Jeans and sonic lengths.

\begin{figure*}
\centering\vspace{-0.5cm}
\includegraphics[scale=0.70]{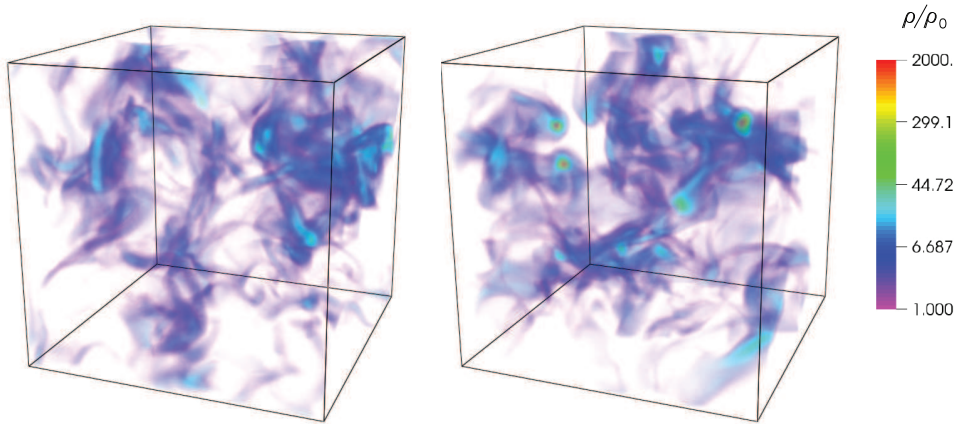}
\vspace{-0.5cm}
\caption{Volume renderings of two density fields at $t/t_{\rm ff}=0.0$ (left) and $t/t_{\rm ff}=1.2$ (right), respectively, where $t_{\rm ff}$ is the free-fall time with an initial density.  The evolutionary time is set to zero when self-gravity is on. Colours in a bar are mapped onto density values normalised with an initial value, $\rho_0$.}
\end{figure*}

The numerical simulations mentioned in the first paragraph didn't include self-gravity of gas.  It is, however, easily expected that the inclusion of self-gravity in a turbulent simulation results in a density PDF with an extended tail at higher densities due to the very nature of self-gravity, which has been, in fact, reported in a few literatures.  \citet{kle00} showed the extended tails of density PDFs from his decay and driven hydrodynamic simulations with self-gravity. \citet{dib05} reported a similar extended tail of a density PDF from one of their numerical models with self-gravity.  \citet{fed08a} showed that power laws develop at the high density tail of the density PDFs of tracer particles in self-gravitating, supersonic turbulence calculations.  Finally, \citet{vaz08} performed self-gravitating, isothermal, hydrodynamic simulations and showed progressive increase of the volume fraction of density PDFs at high densities as a function of time.

The motivation of this paper is to show quantitatively how much CFR$_{\rm ff}$ in a turbulent cloud can be increased by self-gravity with respective to that measured based on a lognormal density PDF.  As we mentioned above, density PDFs from isothermal simulations
without self-gravity result in log-normal functions.  Dense filaments formed in those
simulations are transient, which reside in the high end of the functions and are the main
contributor to core or star formation rates.  It is self-gravity that enables some
of them to develop into collapsing cores, and helps the cores to accrete nearby gas at later
evolutionary stages.  With this motivation in our mind, we perform a numerical simulation for core formation
in a self-gravitating, magnetically supercritical, supersonically turbulent, isothermal cloud.
We show that the volume fractions of density PDFs at higher densities are increasing as time goes on. Even though the increments of the volume fractions at high densities over a lognormal distribution are small, the increments of mass fraction are quite large due to the very high densities.  This, in fact, leads to more significant increment of CFR$_{\rm ff}$ than the one measured based on the lognormal density PDF.

\section{Numerical Method}

In order to see the formation and evolution of cores in a turbulent cloud, we numerically integrate the MHD equations with an isothermal equation of state and the Poisson's equation, using a MHD code based on a total variation diminishing scheme \citep{kim99} and a gravity solver based on the fast Fourier transform. As an initial condition, a uniformly magnetized medium is assumed in a rectangular box.  The periodic boundary condition is imposed in each direction of the box.  The combined isothermal MHD and Poisson's equations are scale-free, and can be written in a dimensionless form with two parameters.  One is the plasma beta, $\beta$, the ratio of gas to magnetic pressures, and the other is the Jeans number, $J$, the ratio of the length of one side of the computational box, $L$, to the initial Jeans length, $\lambda_{J0}=(\pi a_s^2/G\rho_0)^{1/2}$, where $a_s$ is an isothermal sound speed, $G$ is the gravitational constant, and $\rho_0$ is an initial density.

We basically follow the recipes for turbulence generation in \citet{sto98} and \citet{mac99}.  Velocity fluctuations are generated in a Fourier space with the same functional form of the velocity power spectrum in \citet{sto98}.  We, however, take the peak wavenumber, $2(2\pi/L)$, which is smaller than their choice. We then transform the fluctuations into a real space, and adjust their amplitudes in a way that an input kinetic energy rate is a constant.
The level of turbulence driven by this method can be parameterized by a rms (root-mean-square) sonic Mach number, $M_s$.  We start a driven turbulence simulation without turning self-gravity on, and wait until a converged turbulent flow is developed.  Then, we turn self-gravity on and set the time to zero.  The forcing for turbulence generation is still active afterwards.
Numerical simulations based on the above-mentioned method with different parameters have been done.
In this Letter, we show results from one $512^3$ simulation with parameters, $\beta=0.1$, $J=4$, and $M_s=10$.

\section{Results}

To visualise the evolution of density fields as a function of time, the series of volume rendering images of three-dimensional density fields are made.  Two of them are shown in Figure~1. The left image shows a density field at time just before turning self-gravity on, $t/t_{\rm ff}=0$, where $t_{\rm ff}=[3\pi/(32G\rho_0)]^{1/2}$, the free-fall time with an initial or a mean density inside the computational box. The colour bar only covers a normalised density range from 1 to 2000, which enables us to see the distribution of high density gas more clearly.  There are many dense filaments in the left image.  These are formed by the interaction of large-scale supersonic flows with a sonic Mach number around 10.  Isothermal shocks along a field line easily increase the normalised densities at post shock regions up to around 100, which is the square of the Mach number.  After turning self-gravity on, some of the filaments increase their central densities either by merging with each other or accreting nearby gas.  Several condensations are seen in red or yellow colours in the right image at $t/t_{\rm ff}=1.2$, whose normalised central densities are larger than 300.
It is these condensations that are eventually evolved into collapsing cores.  In fact, they are the main contributors to form bumps raised on top of a lognormal density PDF at high densities (see Figure~2).

\begin{figure}
\centering
\includegraphics[scale=0.4]{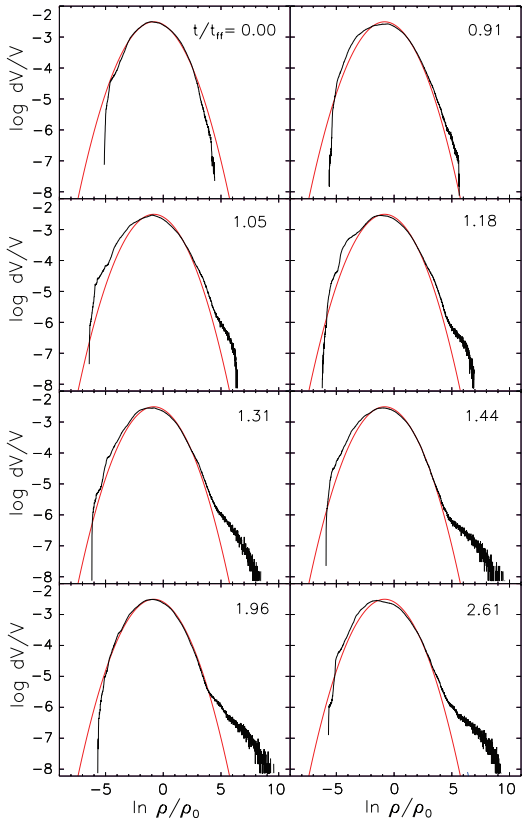}
\caption{Evolution of the volume fractions of density PDFs. The horizontal and vertical axes are logarithmic values based on $e$ and 10, respectively.  A black line in each panel is a density PDF in a computational box at a specified time in units of the initial free-fall time.  A red line is a lognormal distribution, equation~(\ref{eq:LN}), fitted with a density PDF averaged over the saturated stage of turbulence before turning self-gravity on.  The fitted parameters are $\mu =-0.82$ and $\sigma = 1.28$.  The same red line is plotted in each panel.}
\end{figure}

We plot two kinds of density PDFs in black and red lines in each panel of Figure~2.  The black lines are calculated from density fields from our numerical simulation at different times.  A same red line in each panel is a fit of an averaged density PDF over a saturated turbulent state before turning self-gravity on with a lognormal distribution,
\begin{equation}
p_{\rm LN}(s) = \frac{1}{\sqrt {2\pi \sigma^{2}}}
\exp\left[-\frac{(s-\mu)^2}{2\sigma^{2}}\right],
\label{eq:LN}
\end{equation}
where $s=\ln(\rho/\rho_0)$, $\mu$ is the mean, and $\sigma$ is the standard deviation.
Since mass has not added into or subtracted from the computational box during our simulation, total mass inside the box is conserved.  The conservation constraint at the initial state and a state where the density PDF is described with the lognormal distribution, $\rho_0 = \int_{-\infty}^{\infty}\rho p_{\rm LN}(s)ds$, provides us a relation, $\mu=-\sigma^2/2$ \citep{pas98}.  This reduces the two parameters in equation~(\ref{eq:LN}) into one. The average for the red line is taken over from $t/t_{\rm ff}=-1.3$ to $t/t_{\rm ff}=0.0$.  The number of density fields used in the average is 50.  The mean value of the averaged density PDF, $p(s)$, is -0.82 calculated from $\Sigma_i s_i p(s_i)$, where $s_i$ is a discrete value of $s$, and $p(s_i)$ is a volume fraction at $s_i$.  A lognormal distribution with $\sigma=1.28$ calculated from the relation $\sigma=(-2\mu)^{1/2}$ is plotted in each panel as a red line. It serves as a fiducial line to see how much a black line at each panel deviates from the log-normal distribution.  Later, we will also use the $\sigma$ value to estimate CFR$_{\rm ff}$ based on the lognormal distribution.

A black line in the upper left panel at $t/t_{\rm ff}=0$ shows a density PDF at the turn-on time of self-gravity. It is quite well matched with the red line, except at low and high density ends.  The differences may be due to finite numerical resolution and the intermittency effect of the turbulence \citep{kri07, fed10a}.  We then deliberately choose several times based on shapes of density PDFs.  The shape of density PDFs, as time goes on, is hardly changed from the log-normal distribution up to around $t/t_{\rm ff}=0.91$.  It takes for self-gravity to exercise its control over supersonic turbulent flows and then bring significant change in density PDFs.  In fact, it is at $t/t_{\rm ff}=1.05$ that the black line shows quite an excess of volume fractions at a high density tail, which we call a bump on top of the lognormal distribution.  The bump height is increasing rapidly up to $t/t_{\rm ff}=1.44$, and then almost saturated afterwards. Even though the excess volume fractions over the lognormal distribution is small, the excess mass fractions should be quite significant due to high densities at the bump.  It is this excess mass fractions that contribute significantly to core formation rates calculated in the followings.

\begin{figure}
\centering
\includegraphics[scale=0.4]{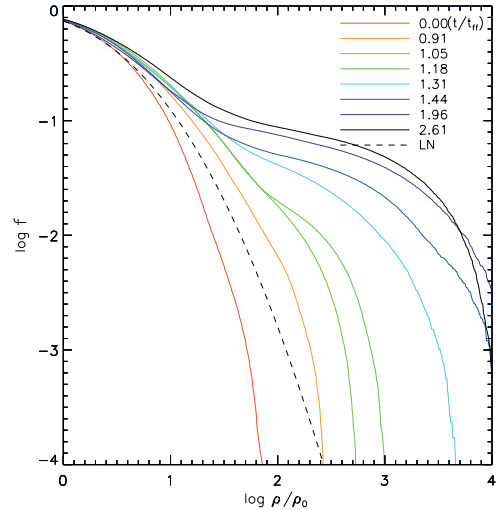}
\caption{Cumulative mass fractions as a function of the normalised density.  Solid lines with different colours are calculated by integrating equation~(\ref{eq:mf}) numerically with the density PDFs at different times shown in Figure~2, respectively.  A dash line is for the lognormal distribution, which is from equation~(\ref{eq:fLN}) with $\sigma=1.28$.}
\end{figure}

We calculate the cumulative mass fractions of the density fields.  Since the volume fraction of a density PDF at each time is known, the cumulative mass fraction, $f(s)$, of gas whose density is larger than $s$ is easily calculated using the following equation,
\begin{equation}
f(s)=\int^{\infty}_{s} \frac{\rho}{\rho_0} p(s)ds = \int^{\infty}_{s} e^sp(s)ds.
\label{eq:mf}
\end{equation}
If $p(s)$ is replaced with equation~(\ref{eq:LN}) with the relation $\mu=-\sigma^2/2$, the integration results in
\begin{equation}
f_{\rm LN}(s) = \frac{1}{2}\left[1+
{\rm erf}\left(\frac{\sigma^2-2s}{2\sqrt{2}\sigma}\right)\right],
\label{eq:fLN}
\end{equation}
which is the same as the expression in equation~(20) for the core formation rate, i.e., ignoring feedback effects, in \citet{kru05}.  We numerically integrate equation~(\ref{eq:mf}) with $p(s)$'s shown in Figure~2, and then plot cumulative mass fraction curves as a function of $\log (\rho/\rho_0)$ in Figure~3.  A dash line is calculated from equation~(\ref{eq:fLN}) with $\sigma=1.28$.  A red line at $t/t_{\rm ff}=0.0$ always lies below the dashed line.  This is due to the fact that the density PDF at $t/t_{\rm ff}=0.0$ drawn with a black line shown in Figure~2 has smaller volume fractions at high densities than the red line.  As time goes on, the cumulative mass fraction increases significantly at high densities. As it is expected from Figure~2, the increasing rate is high during the rapid development of the bump from $t/t_{\rm ff}=1.05$ to 1.44.  Furthermore, the mass fraction curves at later times are significant higher than the one calculated from the lognormal distribution, which is totally due to self-gravity.

\begin{figure}
\centering
\includegraphics[scale=0.4]{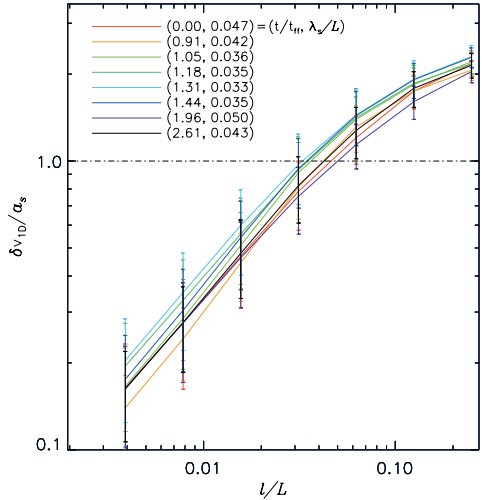}
\caption{One-dimensional velocity dispersion as a function of box size.  Normalization units for the dispersion and the size are the isothermal speed $a_s$ and the one-dimensional size $L$ of the computational box, respectively.  For a given time, measured mean values of velocity dispersions at different box sizes are connected, and the standard deviations are plotted with error bars. The normalised times $t/t_{\rm ff}$ and sonic lengths $\lambda_s/L$ are given in parentheses.}
\end{figure}

The cumulative mass fraction curves shown in Figure~3 may provide CFR$_{\rm ff}$, if a critical density for gas collapse is properly defined.  One idea that has been used is to compare a local Jeans length, $\lambda_J=(\pi a_s^2/G\rho)^{1/2}$, and a sonic length, $\lambda_s$ \citep{pad95, vaz03, kru05}.  In fact, the condition of $\lambda_J=\lambda_s$ gives a  critical density $\rho_c$ normalised with the initial density,
\begin{equation}
\frac{\rho_c}{\rho_0} = \left(\frac{\lambda_{J0}}{\lambda_s}\right)^2,
\label{eq:rhocr}
\end{equation}
where $\lambda_{J0}$ is the initial Jeans length.  The physical background of this condition is that a transonic turbulent velocity dispersion over a volume defined by $\lambda_s$ is barley able to support gravitational collapse of the gas inside the volume.  Since $\lambda_{J0}=L/4$, which is given as an initial condition, we need $\lambda_s$ in term of $L$.  For the calculation of $\lambda_s$, we follow the method in \citet{vaz03}.  We take 100 random positions in the computational box, put a same-sized box less than $L$ centered at each position, and calculate one-dimensional velocity dispersions of turbulent velocity fields in the 100 boxes.  Figure~4 shows the velocity dispersions normalised with the isothermal sound speed as a function of a normalised box size.  The dispersions at eight different times are plotted with coloured solid lines.  Error bars at the measured box size are included, even though they are hardly distinguished from each other.  A horizontal dash-dot line with a unit normalised velocity dispersion is drawn.  In fact, sonic lengths are determined by the meeting points of the dash-dot line and solid lines.  The measured sonic lengths at different times are included in the panel, which are in the range, $0.033L \stackrel{<}{_\sim} \lambda_s \stackrel{<}{_\sim} 0.050L$.   If this range of $\lambda_s$ values and $\lambda_{J0}=L/4$ are plugged in equation~(\ref{eq:rhocr}), then $25 \stackrel{<}{_\sim} \rho_c/\rho_0 \stackrel{<}{_\sim} 57$.  Considering the fact that a density jump brought by a Mach 10 isothermal shock along a magnetic field line is around 100, the range of the normalised critical density can be easily attained by shocks only.  So it is not guaranteed for the gas to collapse, whose normalised density is the upper bound of the critical density range, 57.  Furthermore, the magnetic field included in our simulation plays a certain role in supporting cores.  So the argument based on the local Jeans and sonic lengths may not
give a correct critical density for measuring CFR$_{\rm ff}$ in our numerical simulation.

In order to truly measure CFR$_{\rm ff}$ in our simulation, we first
calculate core formation efficiency, which is the ratio of the total mass of cores defined by a critical density to the total mass in the computational box.  Figure~5 shows the evolution of core formation efficiencies with several different critical densities.  Because of the reason given in the previous paragraph, we choose a rather wide range of critical densities normalised with the initial density, from 30 to 500.  The core formation efficiencies shown in red and brown lines show more or less constant levels up to around $t=0.8t_{\rm ff}$.  There is very tiny fraction of the total mass during the interval, whose normalised density is larger than 100.  This is due to the fact that it takes for self-gravity to exercise its control over turbulent flows, as we have seen in the PDF plots (Figure~2).  After that point, the efficiencies increase rapidly up to around $t=1.5t_{\rm ff}$ and then more slowly later on.  The increase of the efficiencies is mostly due to the accretion of nearby gas onto several cores.

For a given critical density, CFR$_{\rm ff}$ can be calculated using equation~(\ref{eq:fLN}).  Here the free-fall time is again measured based on a mean density of a molecular cloud.  For the
seven normalised critical density values, 30, 50, 100, 200, 300, 400, and 500 shown in Figure~5, the CFR$_{\rm ff}$'s are $2.2\times10^{-2}$, $7.8\times10^{-3}$, $1.5\times10^{-3}$, $2.3\times10^{-4}$, $6.8\times10^{-5}$, $2.7\times10^{-5}$, $1.2\times10^{-5}$, respectively.  In order to measure the CFR$_{\rm ff}$'s from our numerical simulation, we make a least-square fitting of each curve from $t=0$ to
$t=3t_{\rm ff}$ in Figure 5 with a straight line.  In fact, the slopes of those lines give us CFR$_{\rm ff}$'s.  They are 0.048, 0.045, 0.040, 0.036, 0.033, 0.031, 0.029 for normalised critical density values, 30, 50, 100, 200, 300, 400, and 500, respectively. Comparison of theoretical and numerical estimates shows that the CFR$_{\rm ff}$'s from the numerical experiment are 2.2, 5.8, 27, 160, 490, 1100, and 2400 times larger than those from the lognormal distribution for the critical density values 30, 50, 100, 200, 300, 400, and 500, respectively.
The difference becomes larger as the critical density increases.  If we pick 100 up as our fiducial, normalised critical density, the CFR$_{\rm ff}$ based on the theory is likely to be underestimated by about a factor 30.

\begin{figure}
\vspace{-0.5cm}
\centering
\includegraphics[scale=0.4]{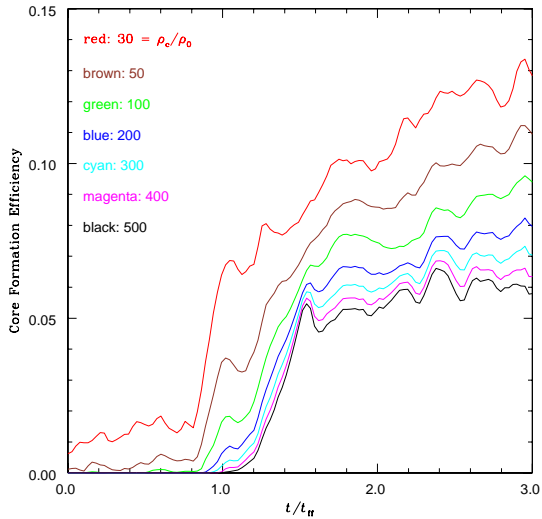}
\vspace{-0.8cm}
\caption{Evolution of core formation efficiencies with different critical densities. The core
formation efficiency is defined as a fraction of total mass in the computational box,
whose density is larger than a critical density.  The time is measured in units of free-fall time with an initial density.  The critical density, $\rho_c$, is normalised with the initial
density, $\rho_0$.}
\end{figure}

\section{Discussions and Conclusions}

Recently \citet{kai09} catalogued column density PDFs of 23 molecular cloud complexes from the 2MASS archive.  They classified them into two groups based on star formation activity and compared their column density PDFs with each other.  The column density PDFs of star-forming clouds always have extended tails, whereas the PDFs of clouds without active star formation follow lognormal distributions or a bit excess at high column densities.  Furthermore, the cumulative fractions of column density PDFs with star-forming clouds are significantly larger than those without active star formation.  These observational results clearly show that self-gravity plays a role in forming the extended tails of the PDFs, which is consistent with our work.

We remind that the extended tails of density PDFs at high densities from isothermal simulations with self-gravity have been shown in a few previous literatures.  The new finding in this Letter is that the extended tails can enhance CFR$_{\rm ff}$ quite significantly.  Cores in a turbulent cloud cannot form without self-gravity.  Core (star) formation rates or core (stellar) mass functions should be measured in the context of a self-gravitating cloud.
Therefore, it is likely that core formation rates previously measured based on the lognormal density PDF \citep{kru05, elm08} are underestimated.  Likewise, the core or stellar initial mass functions based on the lognormal distribution \citep{pad02, hen08} need to be modified.

There are uncertainties in our results due to numerical resolution, measuring CFR$_{\rm ff}$ based on only a density threshold, violation of the numerical Jeans condition, and stellar feedback.  Because of the limited space, we briefly discuss on the last three.  Firstly, the density threshold alone may not fully capture collapsing gas.  More elaborated collapse indicators are needed (for example, \citet{fed10b}). Secondly, Figure~2, for example, shows that density values at a high density tail especially at later stages of our simulation go above the maximum normalised density value, 1024, (see, Equation~(9) in \citet{vaz05}) constrained by the numerical Jeans condition for preventing artificial fragmentation \citep{tru97}. In this Letter we are interested in not the fragmentation of cores but the total amount of mass of cores defined by a critical density. So the violation of the condition will not change our main result but add uncertainty in the measured CFR$_{\rm ff}$.  Thirdly, we didn't include feedback processes from the stars that might form in our simulation. Without the stellar feedback core formation efficiency eventually approaches one. However, at least, before the formation of a first star in the simulation, our
measurement of the core formation efficiency is quite right. In order to
properly measure the core formation efficiency, especially, at the later evolutionary state of a molecular cloud, one should include the feedback.

We performed a magnetically supercritical, supersonic turbulence simulation with the isothermal equation of state to study the effects of self-gravity on density PDFs and the core formation rate.  Here are
conclusions from the study.  First, self-gravity helps to form the extended tail of a density PDF at high densities, which significantly increases CFR$_{\rm ff}$.  Second, the normalised critical density for core collapse determined by the equal condition between the local Jeans and sonic lengths is $25 \stackrel{<}{_\sim} \rho_c/\rho_0 \stackrel{<}{_\sim} 57$, which is smaller than 100, a density jump brought by an isothermal Mach 10 shock in our simulation. So the determined
critical density may not give a correct condition for the core formation.  Third, for our fiducial
normalised critical density, 100, CFR$_{\rm ff}=0.045$, measured from
our numerical simulation is about 30 times larger than
the one, 0.0015, based on the lognormal distribution.
Therefore, self-gravity plays a significant role in enhancing CFR$_{\rm ff}$ in a turbulent cloud.

\section*{Acknowledgments}
The authors thank the referee for constructive comments. The work of J.K. was supported by the Korea Foundation for International Cooperation of Science and Technology through K20702020016-07E0200-01610 and the National Research Foundation of Korea through 2009-0062863 (ARCSEC).

\end{document}